\documentclass[12pt,eadjoint tfnpsf]{article}

\usepackage{amsmath,amsbsy,amssymb,latexsym}

\newcommand{\NN}{\nonumber}

\newcommand{\BE}{\begin{equation}}
\newcommand{\EE}{\end{equation}}
\newcommand{\BEA}{\begin{eqnarray}}
\newcommand{\EEA}{\end{eqnarray}}

\def\12{\frac{1}{2}}
\def\bea{\begin{eqnarray}}
\def\eea{\end{eqnarray}}
\def\ba{\begin{array}}
\def\ea{\end{array}}

\def\one-loop{\mbox{\scriptsize one-loop}}

\def\G{\Gamma}

\catcode`\@=11

\@addtoreset{equation}{section}
\def\theequation{\arabic{section}.\arabic{equation}}

\def\@normalsize{\@setsize\normalsize{15pt}\xiipt\@xiipt
\abovedisplayskip 14pt plus3pt minus3pt%
\belowdisplayskip \abovedisplayskip
\abovedisplayshortskip \z@ plus3pt%
\belowdisplayshortskip 7pt plus3.5pt minus0pt}

\def\small{\@setsize\small{13.6pt}\xipt\@xipt
\abovedisplayskip 13pt plus3pt minus3pt%
\belowdisplayskip \abovedisplayskip
\abovedisplayshortskip \z@ plus3pt%
\belowdisplayshortskip 7pt plus3.5pt minus0pt
\def\@listi{\parsep 4.5pt plus 2pt minus 1pt
\itemsep \parsep
\topsep 9pt plus 3pt minus 3pt}}

\def\underline#1{\relax\ifmmode\@@underline#1\else
$\@@underline{\hbox{#1}}$\relax\fi}
\@twosidetrue

\relax

\catcode`@=12

\catcode`\@=11

\def\section{\@startsection{section}{1}{\z@}{3.5ex plus 1ex minus
.2ex}{2.3ex plus .2ex}{\large\bf}}

\def\thesection{\Roman{section}.}

\def\appendix{\setcounter{section}{0}
\def\thesection{Appendix }
\def\theequation{\Alph{section}.\arabic{equation}}}

\catcode`\@=12

\relax

\def\figcap{\section*{Figure Captions\markboth
{FIGURECAPTIONS}{FIGURECAPTIONS}}\list
{Fig. \arabic{enumi}:\hfill}{\settowidth\labelwidth{Fig. 999:}
\leftmargin\labelwidth
\advance\leftmargin\labelsep\usecounter{enumi}}}
 \relax
\def\tablecap{\section*{Table Captions\markboth
{TABLECAPTIONS}{TABLECAPTIONS}}\list
{Table \arabic{enumi}:\hfill}{\settowidth\labelwidth{Table 999:}
\leftmargin\labelwidth
\advance\leftmargin\labelsep\usecounter{enumi}}}
 \relax
\def\reflist{\section*{References\markboth
{REFLIST}{REFLIST}}\list
{[\arabic{enumi}]\hfill}{\settowidth\labelwidth{[999]}
\leftmargin\labelwidth
\advance\leftmargin\labelsep\usecounter{enumi}}}
 \relax

\newskip\humongous \humongous=0pt plus 1000pt minus 1000pt

\newif\ifdtup

\def\Im{\mathop{\rm Im}}
\def\Re{\mathop{\rm Re}}

\def\beq{\begin{equation}}
\def\eeq{\end{equation}}

\def\beqn{\begin{eqnarray}}
\def\eeqn{\end{eqnarray}}
\relax

\def\G2{{\; \rm GeV/}c2}
\def\G{\; \rm GeV}

\def\dotx{\dotx{\dot\overline{x}}}

\relax

\newcommand\CD{{\mathcal D}}
\newcommand\CF{{\mathcal F}}

\newcommand\CK{{\mathcal K}}
\newcommand\CL{{\mathcal L}}

\newcommand\CN{{\mathcal N}}

\newcommand\CS{{\mathcal S}}

\newcommand\CW{{\mathcal W}}

\newcommand\I{{\mathbb I}}

\def\LLangle{\left< \hspace{-2.5mm} \left<}
\def\RRangle{\right> \hspace{-2.5mm} \right>}

\def\llangle{\left< \hspace{-1mm} \left<}
\def\rrangle{\right> \hspace{-1mm} \right>}

\renewcommand{\thefootnote}{\fnsymbol{footnote}}
\begin{document}
%
%
\begin{titlepage}

\begin{flushright}
\normalsize
~~~~
March, 2006 \\
OCU-PHYS 243 \\
hep-th/0603180 \\
\end{flushright}

%
\begin{center}
{\Large\bf   $U(N)$ Gauged  $\mathcal{N}=2$ Supergravity and  \\
 Partial Breaking of Local ${\cal N}=2$ Supersymmetry }
\end{center}

\vfill

\begin{center}
{%
H. Itoyama$^{a,b}$\footnote{e-mail: itoyama@sci.osaka-cu.ac.jp}
\quad and \quad
K. Maruyoshi$^a$\footnote{e-mail: maruchan@sci.osaka-cu.ac.jp}
}
\end{center}

\vfill

\begin{center}
$^a$ \it Department of Mathematics and Physics,
Graduate School of Science\\
Osaka City University\\
\medskip

$^b$ \it Osaka City University Advanced Mathematical Institute
(OCAMI)

\bigskip

3-3-138, Sugimoto, Sumiyoshi-ku, Osaka, 558-8585, Japan \\

\end{center}

\vfill

\begin{abstract}
  We study a $U(N)$ gauged $\mathcal{N}=2$ supergravity model
  with one hypermultiplet parametrizing $SO(4,1)/SO(4)$ quaternionic manifold.
  Local $\mathcal{N}=2$ supersymmetry is known to be spontaneously broken
  to $\mathcal{N}=1$ in the Higgs phase of $U(1)_{graviphoton} \times U(1)$. 
  Several properties are obtained of this model in the vacuum of unbroken $SU(N)$ gauge group. 
  In particular, we derive mass spectrum analogous to the rigid counterpart 
  and put the entire resulting potential on this vacuum in the standard superpotential form 
  of $\mathcal{N}=1$ supergravity.

\end{abstract}

\vfill

\setcounter{footnote}{0}
\renewcommand{\thefootnote}{\arabic{footnote}}

\end{titlepage}

\section{Introduction}

  For more than a decade, $\mathcal{N}=2$ supersymmetry both in its local and rigid realizations has played an
  important role in the theoretical developments of quantum field theory and particle physics.
  It has led us to the subject of exactly determined low energy effective actions 
  \cite{SeibergWitten1, SeibergWitten2} and has inspired the construction of Lagrangians based on special
  K\"ahler geometry \cite{Castellani, Strominger, D'Auria, Andrianopoli0, Andrianopoli, VanPro, VanPro2, VanPro3}. 
  These achievements have proven valuable in order to analyze some of the phenomena which occur in string theory. 

  Spontaneous breaking of $\mathcal{N}=2$ supersymmetry to $\mathcal{N}=1$ is an interesting problem 
  in the light of its implications of string theory to the low energy $\mathcal{N}=1$ supersymmety, 
  which is phenomenologically promising.
  We give here a partial list of the references of this subject on the linear realizations 
  \cite{Cecotti2, Ferrara1, Ferrara2, APT, APT2, APT3, Fre, Porrati, Louis, Andrianopoli1, Andrianopoli2, FIS12, FIS13, FIS3, FIS4}.
  In particular, spontaneous partial breaking of rigid $\mathcal{N}=2$ supersymmetry in the $U(N)$ gauge model 
  with or without hypermultiplets has been demonstrated recently \cite{FIS12, FIS13, FIS3, FIS4} 
  under a generic breaking pattern of  the $U(N)$ gauge symmetry.
  Several other properties of this model have been obtained. 
  It should be emphasized that the partial breaking of rigid $\mathcal{N}=2$ supersymmetry
  is realized here in the Coulomb phase of overall $U(1)$,
  the Nambu-Goldstone fermion being the superpartner of
  the massless photon and  that both interact with the $SU(N)$ sector
  thanks to the non-Lie algebraic property of the prepotential.

  There are already considerable differences between the rigid special K\"ahler geometry and 
  its local counterpart and between the hyperk\"ahler geometry
  and the quaternionic geometry as have been emphasized in the
  literatures \cite{Andrianopoli, VanPro, VanPro2}.  
  This is bound to be reflected in the comparison of the vacuum analysis
  of a rigid $\mathcal{N}=2$ effective action with its supergravity counterpart. 
  This will be a thrust of the present paper.
  Spontaneous partial breaking of local $\mathcal{N}=2$ supersymmetry has been studied in 
   \cite{Cecotti2, Ferrara1, Ferrara2, Fre, Porrati, Louis, Andrianopoli1, Andrianopoli2}. 
  It was noted from the beginning that both the Higgs and the super-Higgs mechanisms must take place simultaneously 
  and that the vacuum must lie in the Higgs phase of $U(1)_{graviphoton} \times U(1)$.
  The tight structure of the spectrum produced by the mechanisms requires
  at least one hypermultiplet with two $ U(1) $ translational isometries to be introduced in the models.

  In this paper, we study some of the basic and yet unexplored properties
  of the $U(N)$ gauged $\mathcal{N}=2$ supergravity model
  in which local $\mathcal{N}=2$ supersymmetry is partially broken spontaneously.
  In particular, the holomorphic section of our model is chosen as a generic function 
  which leads to the nontrivial scalar coupling terms and the scalar potential.
  (In reference \cite{Ferrara1}, a simple form of the section has been adopted.)
  In the next section, we briefly review  $U(N)$ gauged $\mathcal{N}=2$ supergravity in four dimensions 
  and consider the model which contains a $U(N)$ vector multiplet and
  a hypermultiplet parametrizing $SO(4,1)/SO(4)$ quaternionic manifold.
  Because of the choice of the section we need careful consideration of the vacuum, which is done in section three.
  We consider and solve the vacuum conditions of the model under the assumption of unbroken $SU(N)$ gauge symmetry. 
  The second vacuum condition, which is a variation of the potential with respect to the hypermultiplet scalar $b^u$, 
  was not considered before and the super-Higgs mechanism can not operate without this one.
  Partial breaking of local $\mathcal{N}=2$ supersymmetry is exhibited.
  In section four, we derive the mass spectrum of the model
  and interpret it in terms of  $\mathcal{N}=1$ on-shell supermultiplets.
  In section five, we construct the entire Lagrangian on this vacuum and express it in terms of two superpotentials
  which are related to each other by a simple relation (\ref{superpotential}). 
  The resulting form conforms to the standard form of $\mathcal{N}=1$ supergravity.

\section{$ U(N) $ Gauged  $\mathcal{N}=2$ Supergravity}
  The field contents of  $ U(N) $ Gauged  $\mathcal{N}=2$ Supergravity
  are summarized as follows:
    \begin{itemize}
      \item \textit{supergravity multiplet}
      \\
       consisting of the vierbein $ e^i_\mu $ ($i,\mu = 0,1,2,3$), 
      two gravitini $ \psi^A_\mu $ ($A=1,2$) and the graviphoton $ A^0_\mu $.
      (The upper and the lower position of the index $ A $ represent left and right chirality respectively.)
      \item \textit{vector multiplet}
      \\
      consisting of a gauge boson $ A^a_\mu $, 
      two gaugini $ \lambda^{aA} $ and a complex scalar $ z^a $.
      The index $ a $   ($ a = 1,\dots,N^2 $) labels the generators
      of the $ U(N) $ gauge group and $a= N^2 \equiv n $ refers to
      the overall $U(1)$. (The notation on the chirality is opposite to that of  the gravitini, namely, 
      the upper and the lower position denote right and left chirality respectively.)
      \item \textit{hypermultiplet}
      \\
      consisting of two hyperini $ \zeta^\alpha $ ($ \alpha = 1,2 $) 
      and four real scalars $ b^u $ ($ u = 0,1,2,3 $).
      (The upper and the lower position of the index $ \alpha $ represent left and right chirality respectively.)
    \end{itemize}
  General construction of the Lagrangian of gauged $\mathcal{N}=2$
  supergravity has been given in \cite{Castellani, D'Auria, Andrianopoli}.
  We exhibit  the parts of
  the Lagrangian and the supersymmetric transformation laws 
  which are necessary for our analysis of the vacuum.
  
\subsection{Vector Multiplet}
  The manifold associated with the vector multiplet is
  special K\"ahler of the local type 
  \cite{Castellani, Strominger, D'Auria, Andrianopoli0, Andrianopoli, VanPro, VanPro2, VanPro3}.
  It is equipped with a holomorphic section, 
    \begin{equation}
      \Omega (z)
        =   \left(\begin{array}{cc}
            X^\Lambda (z) \\
            F_\Lambda (z) \\
            \end{array}
            \right),~~~~
            \Lambda = 0, 1, \ldots, n 
    \end{equation}
  The index 0 refers to the graviphoton part.
  In terms of this section, the K\"ahler potential is given by,
    \BE
      \mathcal{K} 
        = - \log i \left\langle\Omega | \bar\Omega\right\rangle
        = - \log i ( \bar{X}^\Lambda F_\Lambda - X^\Lambda \bar{F}_\Lambda ),
    \EE
  where
    \begin{equation}
      i \left\langle\Omega | \bar\Omega\right\rangle
        \equiv 
          - i \Omega^T
            \left(
            \begin{array}{cc}
            0 & \I \\
          - \I & 0 \\
            \end{array}
            \right) \Omega^*.
    \end{equation}
  The non-holomorphic section is introduced by
    \begin{equation}
      V 
       =    \left(\begin{array}{cc}
            L^\Lambda \\
            M_\Lambda \\
            \end{array}
            \right)
       \equiv
            e^{\CK/2} \Omega
       =    e^{\CK/2} 
            \left(\begin{array}{cc}
            X^\Lambda \\
            F_\Lambda \\
            \end{array}
            \right),
    \end{equation}
  and its covariant derivative is
    \begin{equation}
      U_a
       \equiv 
            \nabla_a V
       =    (\partial_a + \frac{1}{2} \partial_a \mathcal{K} ) V
       \equiv 
            \left(\begin{array}{cc}
            f^\Lambda_a \\
            h_{\Lambda \vert a} \\
            \end{array}
            \right).
    \end{equation}
  One characteristic property of $\mathcal{N}=2$ supergravity,
  which follows from the special  K\"ahler geometry of the local type,
  is the existence of  totally symmetric rank-three tensor
  $ C_{abc} $ such that
    \begin{equation}
      \nabla_a U_b 
       =    i C_{abc} g^{cd^*} \bar{U}_{d^*}.
    \end{equation}
  (See, for example,  \cite{Andrianopoli, VanPro}). 
  The generalized gauge coupling matrix  $\bar{\mathcal{N}}_{\Lambda \Sigma}$
  is introduced via the following relations:
    \begin{equation}
      \bar{M}_\Lambda 
       =    \bar{\mathcal{N}}_{\Lambda \Sigma} \bar{L}^\Sigma,~~~~
      h_{\Lambda \vert a}
       =    \bar{\mathcal{N}}_{\Lambda \Sigma} f^\Sigma_a .
    \end{equation}
  The solution is given in terms of  two $(n+1) \times (n+1)$ matrices
    \begin{equation}
      f^\Lambda_I
       =    \left(\begin{array}{cc}
            f^\Lambda_a \\
            \bar{L}^\Lambda \\
            \end{array}
            \right),~~~~
     h_{\Lambda \vert I}
      =     \left(\begin{array}{cc}
            h_{\Lambda \vert a} \\
            \bar{M}_\Lambda \\
            \end{array}
            \right)
    \end{equation}
  as 
    \begin{equation}
      \bar{\mathcal{N}}_{\Lambda \Sigma}
       =    h_{\Lambda \vert I}   (f^{-1})^I_\Sigma.
            \label{CouplingMatrix}
    \end{equation}
  It is well-known that this quantity appears in the kinetic term of
  the gauge bosons.
  
  To specify the model, we need to choose the holomorphic section.
  Our choice, which is essentially  that of \cite{Ferrara2}, is 
    \begin{eqnarray}
      X^0(z)
       =    \frac{1}{\sqrt{2}},~~~~
      F_0(z)
      &=&   \frac{1}{\sqrt{2}} \left( 2 \CF(z) - z^a \frac{\partial \CF(z)}{\partial z^a} \right),
            \nonumber \\  
      X^{\hat{a}}(z)
       =    \frac{1}{\sqrt{2}} z^{\hat{a}},~~~~
      F_{\hat{a}}(z)
      &=&   \frac{1}{\sqrt{2}} \frac{\partial \CF(z)}{\partial z^{\hat{a}}},
            \\ \label{prepotential}
      X^n(z)
       =    \frac{1}{\sqrt{2}} \frac{\partial \CF(z)}{\partial z^n},~~~~
      F_n(z)
      &=& - \frac{1}{\sqrt{2}} z^n,
            \nonumber 
    \end{eqnarray}
  where the index $\hat{a} = 1,\ldots, n - 1, $ labels the generators of $SU(N)$ subgroup.
  It has been obtained from the derivatives of the holomorphic function $ F(X^0,X^a) = (X^0)^2 \CF (X^a/X^0) $, 
  that is, $ \partial F / \partial X^\Lambda $ and performing the symplectic transformation 
  $ X^n \rightarrow - F_n, F_n \rightarrow X^n $.
  The K\"ahler potential and its derivatives are given by
    \begin{eqnarray}
      \mathcal{K}
      &=& - \log \mathcal{K}_0,
            \\  
      \partial_a \mathcal{K}_{0}
      &=&   \frac{i}{2} 
            \left(
            \CF_a - \bar{\CF}_a - (z^c - \bar{z}^c) \CF_{ac} 
            \right),
            \\
      g_{ab^*}
      &=&   \partial_a \partial_{b^*} \mathcal{K}
            \nonumber \\
      &=&   \partial_a \CK \partial_{b^*} \CK - \frac{i}{2 \CK_0} (\CF_{ab} - \bar{\CF}_{ab}),
            \label{gab}
    \end{eqnarray}
  where $ \CF_a = \partial \CF/ \partial z^a $ and
    \begin{equation}
      \mathcal{K}_0 
       =    i 
            \left(
            \CF - \bar{\CF} - \frac{1}{2} ( z^a - \bar{z}^a ) (\CF_a + \bar{\CF}_a) 
            \right).
    \end{equation}
  Furthermore, the covariant derivative of $ f^\Lambda_a $ is
    \begin{eqnarray}
      \nabla_a f^0_b
      &\equiv&    
            \frac{i e^{\mathcal{K}/2}}{\sqrt{2}} C_{abc} g^{cd^*} \partial_{d^*} \mathcal{K}
            \nonumber \\
      &\equiv& 
            \partial_a f^0_b 
          + \Gamma^c_{ab} f^0_c
          + \frac{1}{2} \partial_a \mathcal{K} f^0_b
            \nonumber \\
      &=&   \frac{e^{\mathcal{K}/2}}{\sqrt{2}} 
            \left(
            \partial_a \partial_b \mathcal{K}
          - \partial_a \CK \partial_b \mathcal{K}
          + \frac{1}{\mathcal{K}_0} g^{cd^*} (\partial_a \partial_b \mathcal{K}_0 \partial_{d^*} \mathcal{K}
          + \partial_a \partial_b \partial_{d^*} \mathcal{K}_0) \partial_c \mathcal{K}
            \right)
            \label{cdf0},
            \\
      \nabla_a f^n_b
      &\equiv&    
            \frac{i e^{\mathcal{K}/2}}{\sqrt{2}} C_{abc} g^{cd^*} 
            (\bar{\CF}_{nd} + \partial_{d^*} \mathcal{K} \bar{\CF}_n)
            \nonumber \\
      &=&   \frac{e^{\mathcal{K}/2}}{\sqrt{2}} 
            \left(
            \CF_{nab}
          - \partial_a \mathcal{K} \partial_b \mathcal{K} \CF_n
          + \partial_a \partial_b \mathcal{K} \CF_n
            \right)
            \nonumber \\
      & & + \frac{e^{\mathcal{K}/2}}{\sqrt{2} \mathcal{K}_0} g^{cd^*} 
            \left(
            \partial_a \partial_b \mathcal{K}_0 \partial_{d^*} \mathcal{K}
          + \partial_a \partial_b \partial_{d^*} \mathcal{K}_0
            \right) 
            (\CF_{nc} + \partial_c \mathcal{K} \CF_n).
            \label{cdfn}
    \end{eqnarray}
  The Christoffel symbol is defined as $ \Gamma^c_{ab} = - g^{cd^*} \partial_b g_{ad^*} $.
  These equations will be used in the analysis of the potential term.

  In order gauge the vector multiplet, first introduce the Killing vectors  which are defined by
    \begin{eqnarray}
      k^c_a \partial_c 
       =   f^c_{ab} z^b \partial_c,~~~~
      k^{c^*}_a \bar{\partial}_{c^*}
       =   f^c_{ab} \bar{z}^{b^*} \bar{\partial}_{c^*}, 
           \label{UNgauging}
    \end{eqnarray}
  where $ f^a_{bc} $ is the structure constant of the $ U(N) $ gauge group satisfying
    \begin{equation}
      [t_a, t_b]
        =   i f^c_{ab} t_c.
    \end{equation}
  We will deal with the case in which  the Lie derivative $\mathcal{L}_{\Lambda}$  satisfies
    \BE
      0
       =    \mathcal{L}_{\Lambda} \CK
       =    k^b_{\Lambda} \partial_b \CK
          + k^{b^*}_{\Lambda} \partial_{b^*} \CK.
            \label{killing}
    \EE
  The covariant derivative of the scalar fields, for example, takes the standard form:
    \begin{eqnarray}
      \nabla_\mu z^a
      &=&   \partial_\mu z^a + A^\Lambda_\mu k^a_\Lambda
            \nonumber \\
      &=&   \partial_\mu z^a + f^a_{bc} A^b_\mu z^c.
    \end{eqnarray}
  
\subsection{Hypermultiplet}  
  Four real scalar components $ b^u $ of the hypermultiplet span
  the quaternionic manifold which is taken to be $SO(4,1)/SO(4)$.
  The quaternionic geometry is in general determined by a triplet of quaternionic potentials,
    \begin{eqnarray}
      \Omega^x
      &=&   \Omega^x_{uv} db^u \wedge db^v 
            \\ \nonumber
      &=&   d \omega^x + \frac{1}{2} \epsilon^{xyz} \omega^y \wedge  \omega^z,
            ~~~~ x = 1,2,3,
            \label{Omega}
    \end{eqnarray}
  where $\omega^x = \omega^x_u db^u $ are the $ SU(2) $ connections.
  In this paper, we take the same parametrizations as that of \cite{Ferrara1, Ferrara2}.
  The above quantities read
    \begin{equation}
      \omega^x_u 
       =    \frac{1}{b^0} \delta^x_u,~~~~
      \Omega^x_{0u}
       =  - \frac{1}{2(b^0)^2} \delta^x_u,~~~~
      \Omega^x_{yz}
       =    \frac{1}{2(b^0)^2} \epsilon^{xyz}. 
    \end{equation}
  The metric $ h_{uv} $ of this manifold is
    \begin{equation}
      h_{uv}
       =    \frac{1}{2(b^0)^2} \delta_{uv},
    \end{equation}
  while the symplectic vielbein is   
    \beqn
  \mathcal{U}^{\alpha A}_u db^u,  ( \alpha, A = 1,2),\;\;
    \mathcal{U}^{\alpha A}
       =    \frac{1}{2b^0} \epsilon^{\alpha \beta} (db^0 - i \sigma^x db^x)_\beta^{~A},
    \eeqn
  where $ \sigma^x $ are the standard Pauli matrices.
  
 Let us introduce  the Killing vectors $ k^u_\Lambda $ and the momentum maps $ \mathcal{P}^x_\Lambda $ 
 associated with  two $ U(1) $ translational isometries of this quaternionic manifold \cite{Ferrara2}:
    \begin{eqnarray}
      k^u_0 
      &=&    g_1 \delta^{u3} + g_2 \delta^{u2},~~~~
      k^u_{\hat{a}}
       =     0,~~~~
      k^u_n
       =     g_3 \delta^{u2},
             \nonumber \\
      \mathcal{P}^x_0 
      &=&    \frac{1}{b^0} (g_1 \delta^{x3} + g_2 \delta^{x2}),~~~~
      \mathcal{P}^x_{\hat{a}}
       =     0,~~~~
      \mathcal{P}^x_n
       =     \frac{1}{b^0} g_3 \delta^{x2}.
             \label{hypergauging}
    \end{eqnarray}
  Here $ g_1, g_2, g_3, \in \mathbb{R}$ are coupling constants.
  These constants play the same role as   the superpotential and
 the Fayet-Iliopoulos term do in the rigid theory \cite{APT, FIS12, FIS3, FIS4}.

\subsection{The Lagrangian of $\mathcal{N}=2$ Supergravity}

 Let us write the parts of the Lagrangian of
 the $\mathcal{N}=2$ gauged supergravity which is needed in our
 analysis:
    \begin{equation}
      \CL_{\rm}
       =     \sqrt{- g} 
             \left(
             \CL_{\rm{kin}} + \CL_{\rm{mass}} - V(z,\bar{z},b) + \ldots
             \right),
    \end{equation}
  where
    \begin{eqnarray}
      \CL_{\rm{kin}}
      &=&    R + g_{ab^*} \nabla_{\mu} z^a \nabla^\mu \bar{z}^{b^*}
           + h_{uv} \nabla_\mu b^u \nabla^\mu b^v 
           + \frac{\epsilon^{\mu \nu \lambda \sigma}}{\sqrt{-g}}
             (\bar{\psi}^A_\mu \gamma_\nu \nabla_\lambda \psi_{A \sigma}
           - \bar{\psi}_{A \mu} \gamma_\nu \nabla_\lambda \psi^A_\sigma)
             \nonumber \\
      & &  + \frac{1}{4} (\Im \mathcal{N})_{\Lambda \Sigma} F^\Lambda_{\mu \nu} F^{\Sigma \mu \nu}
           + \frac{1}{4} (\Re \mathcal{N})_{\Lambda \Sigma} F^\Lambda _{\mu \nu} \widetilde{F}^{\Sigma \mu \nu} 
             \nonumber \\
      & &  - i g_{ab^*} \bar{\lambda}^{aA} \gamma_\mu \nabla^\mu \lambda^{b^*}_A 
           - 2 i \bar{\zeta}^{\alpha} \gamma_\mu \nabla^\mu \zeta_\alpha
           + \ldots,
             \\
      \CL_{\rm{Yukawa}} 
      &=&   2 S_{AB} \bar{\psi}^A_\mu \gamma^{\mu \nu} \psi^B_\nu 
          + i g_{ab^*} W^{aAB} \bar{\lambda}^{b^*}_A \gamma_\mu \psi^\mu_B
          + 2 i N^A_\alpha \bar{\zeta}^\alpha \gamma_\mu \psi^\mu_A
            \nonumber \\
      & & + \mathcal{M}^{\alpha \beta} \bar{\zeta}_\alpha \zeta_\beta
          + \mathcal{M}^\alpha_{aB} \bar{\zeta}_\alpha \lambda^{aB}
          + \mathcal{M}_{aA \vert bB} \bar{\lambda}^{a}_A \lambda^{bB}
          + h.c.,   \label{Yukawa}
            \\
      V(z,\bar{z},b)
      &=&   g_{ab^*} k^a_\Lambda k^{b^*}_\Sigma \bar{L}^\Lambda L^\Sigma
          + g^{ab^*} f^\Lambda_a \bar{f}^\Sigma_{b^*} \mathcal{P}^x_\Lambda \mathcal{P}^x_\Sigma
          + 4~ h_{uv} k^u_\Lambda k^v_\Sigma \bar{L}^\Lambda L^\Sigma
          - 3~ \bar{L}^\Lambda L^\Sigma \mathcal{P}^x_\Lambda \mathcal{P}^x_\Sigma.
      \end{eqnarray}
  Here $ F^\Lambda_{\mu \nu} $ are the field strengths of the $ U(N) $ gauge fields and  that of the graviphoton field, 
  and $ \tilde{F}^\Lambda_{\mu \nu} $ are their Hodge duals.
  The supersymmetry transformation laws of the fermions are
    \begin{eqnarray}
      \delta \psi_{A \mu}
      &=&   i S_{AB} \gamma_\mu \epsilon^B
          + \ldots,  
            \label{strns1}
            \\
      \delta \lambda^{aA}
      &=&   W^{aAB} \epsilon_B
          + \ldots,  
            \label{strns2}
            \\
      \delta \zeta_\alpha
      &=&   N^A_\alpha \epsilon_A
          + \ldots.  
            \label{strns3}
    \EEA
 The matrices appearing in the supersymmetry transformation laws
 and  in eq. (\ref{Yukawa}) are  composed of  the geometric quantities listed in the last two subsections:
    \begin{eqnarray}
      S_{AB}
      &=&   \frac{i}{2} (\sigma_x)_{AB} \mathcal{P}^x_\Lambda L^\Lambda,
            \\
      W^{aAB}
      &=&   \epsilon^{AB} k^a_\Lambda \bar{L}^\Lambda
    + i (\sigma_x)^{AB} \mathcal{P}^x_\Lambda g^{ab^*} \bar{f}^\Lambda_{b^*}  \equiv   W^{aAB}_1 +  W^{aAB}_2
            \\
      N^A_\alpha
      &=&   2~ \mathcal{U}^A_{\alpha u} k^u_\Lambda \bar{L}^\Lambda,
            \\
      \mathcal{M}^{\alpha \beta}
      &=& - \mathcal{U}^{A \alpha}_u \mathcal{U}^{B \beta}_v \epsilon_{AB} \nabla^{[u} k^{v]}_\Lambda L^\Lambda,
            \\
      \mathcal{M}^\alpha_{bB}
      &=& - 4~ \mathcal{U}^\alpha_{Bu} k^u_\Lambda f^\Lambda_b,
            \\
      \mathcal{M}_{aA \vert bB}
      &=&   \frac{1}{2} 
            \left( 
            \epsilon_{AB} g_{ac^*} k^{c^*}_\Lambda f^\Lambda_b
       + i (\sigma_x)_{AB} \mathcal{P}^x_\Lambda \nabla_b f^\Lambda_a
            \right)  \\
      &\equiv& 
            \mathcal{M}_{aA \vert bB}^1
         +  \mathcal{M}_{aA \vert bB}^2.
    \end{eqnarray}
  We obtain explicit forms of these matrices from 
  (\ref{prepotential})-(\ref{gab}), (\ref{UNgauging}) and (\ref{Omega})-(\ref{hypergauging}):
    \BEA
      S_{AB}
      &=& - \frac{i e^{\CK/2}}{2 \sqrt{2} b^0} 
            \left(
            \begin{array}{cc}
            i (g_2 +g_3 \CF_n) & g_1 \\
            g_1 & i (g_2 +g_3 \CF_n)
            \end{array}
            \right),
            \label{S} \\
      W^{aAB}_1 
      &=& - i e^{\CK/2} \CD^a
            \left(
            \begin{array}{cc}
            0 & 1 \\
            -1 & 0
            \end{array}
            \right),
            \\
      W^{aAB}_2
      &=&   \frac{e^{\CK/2}}{\sqrt{2} b^0} g^{ab^*}
            \left(
            \begin{array}{cc}
            \displaystyle\scriptstyle{g_2 \partial_{b^*} \CK + g_3 (\bar{\CF}_{nb} + \partial_{b^*} \CK \bar{\CF}_n)} 
            & i g_1 \partial_{b^*} \CK
            \\
            i g_1 \partial_{b^*} \CK
            & 
            \displaystyle\scriptstyle{g_2 \partial_{b^*} \CK + g_3
            (\bar{\CF}_{nb} + \partial_{b^*} \CK \bar{\CF}_n)}
            \end{array}
            \right),
            \\
      N^A_\alpha
      &=&   \frac{i e^{\CK/2}}{\sqrt{2} b^0} 
            \left(
            \begin{array}{cc}
            g_1 & - i (g_2 +g_3 \bar{\CF}_n) \\
            i (g_2 +g_3 \bar{\CF}_n) & - g_1
            \end{array}
            \right),
            \label{N} \\
      \mathcal{M}^{\alpha \beta}
      &=&   \frac{i e^{\CK/2}}{\sqrt{2} b^0}
            \left(
            \begin{array}{cc}
            - i (g_2 +g_3 \CF_n) & g_1 \\
            g_1 & - i (g_2 +g_3 \CF_n)
            \end{array}
            \right),
            \\
      \mathcal{M}^\alpha_{bB}
      &=& - \frac{\sqrt{2} i e^{\CK/2}}{b^0}
            \left(
            \begin{array}{cc}
            g_1 \partial_{a} \CK &
            \displaystyle\scriptstyle{i (g_2 \partial_{a} \CK + g_3 (\CF_{na} + \partial_{a} \CK \CF_n))} 
            \\
            \displaystyle\scriptstyle{- i (g_2 \partial_{a} \CK + g_3 (\CF_{nb} + \partial_a \CK \CF_n))}
            & g_1 \partial_{a} \CK
            \end{array}
            \right),
            \\
      \mathcal{M}_{1;aA \vert bB}
      &=& - \frac{i e^{\CK/2}}{2} g_{ac^*} (\partial_b + \partial_b \CK) \CD^c
            \left(
            \begin{array}{cc}
            0 & 1 \\
            -1 & 0
            \end{array}
            \right),
    \EEA
    \BEA
      \mathcal{M}_{2;aA \vert bB}
      &=& - \frac{1}{2 b^0} 
            \left(
            \begin{array}{cc}
            \displaystyle\scriptstyle{g_2 \nabla_b f_a^0 + g_3 \nabla_b f_a^n} 
            & - i g_1 \nabla_b f_a^0
            \\
            - i g_1 \nabla_b f_a^0
            & 
            \displaystyle\scriptstyle{g_2 \nabla_b f_a^0 + g_3 \nabla_b f_a^n}
            \end{array}
            \right)
            \NN \\
      &=&   \frac{i e^{\CK/2}}{2 \sqrt{2}} C_{abc} g^{cd^*} 
            \left(
            \begin{array}{cc}
            \displaystyle\scriptstyle{g_2 \partial_{b^*} \CK + g_3 (\bar{\CF}_{nb} + \partial_{b^*} \CK \bar{\CF}_n)} 
            & - i g_1 \partial_{b^*} \CK
            \\
            - i g_1 \partial_{b^*} \CK
            & 
            \displaystyle\scriptstyle{g_2 \partial_{b^*} \CK + g_3 (\bar{\CF}_{nb} + \partial_{b^*} \CK \bar{\CF}_n)}
            \end{array}
            \right).
    \EEA
  Here  we have introduced 
    \BEA
      \CD^a
      &=&   \frac{i}{\sqrt{2}} f^a_{~bc} \bar{z}^{b^*} z^c.
    \EEA

\section{Partial Breaking of $\mathcal{N}=2$ Local Supersymmetry}
  By the gauging of hypermultiplet, the scalar potential takes a nontrivial form and is given by 
    \begin{eqnarray}
      V(z,\bar{z},b)
      &=&    e^{\CK} g_{ab^*} \CD^a \CD^b
           + \frac{e^{\mathcal{K}}}{(b^0)^2} g^{ab^*} D^x_a \bar{D}^x_{b^*}
             \nonumber \\
      & &  - \frac{e^{\mathcal{K}}}{2 (b^0)^2} 
             (\mathcal{E}^x + \mathcal{M}^x \CF_n)(\mathcal{E}^x + \mathcal{M}^x \bar{\CF}_n),
             \label{V}
    \end{eqnarray}
  with
    \begin{eqnarray}
      D^x_a
      &=&    \frac{1}{\sqrt{2}} 
             \left(
             \mathcal{E}^x \partial_a \CK + \mathcal{M}^x (\CF_{na} + \partial_a \CK \CF_n)
             \right),
             \\
      \mathcal{E}^x
      &=&    (0,\ g_2,\ g_1),
             \nonumber \\
      \mathcal{M}^x
      &=&    (0,\ g_3,\ 0).
             \nonumber
    \end{eqnarray}
  The first term comes from  the $U(N)$ gauging of the vector multiplet
  while the second and the last terms correspond to gauging of
  the hypermultiplet.
  
  Let us find the conditions which determine the minimum of the potential.
  Let us  first consider the variations of $ V $ with respect to $ z^a $.
  The derivative of  the second and  the third terms of $ V $ reads
    \BEA 
      \lefteqn{\frac{e^{\mathcal{K}}}{(b^0)^2} 
             \left(
             (\partial_a \mathcal{K}) g^{bc^*} D^x_b \bar{D}^x_{c^*}
           + (\partial_a g^{bc^*}) D^x_b \bar{D}^x_{c^*}
           + g^{bc^*} (\partial_a D^x_b) \bar{D}^x_{c^*}
             \right)}
             \nonumber \\
      &=&    \frac{e^{\mathcal{K}}}{(b^0)^2} g^{bc^*} \bar{D}^x_{c^*}
             \left(
             \partial_a D^x_b
           - (\partial_b \mathcal{K}) D^x_a
           + \frac{1}{\mathcal{K}_0} g^{ed^*} (\partial_a \partial_b \mathcal{K}_0 \partial_{d^*} \mathcal{K}
           + \partial_a \partial_b \partial_{d^*} \mathcal{K}_0) D^x_e
             \right)
             \nonumber \\
      &=&    \frac{i e^{\mathcal{K}}}{(b^0)^2} C_{abc} g^{bd^*} \bar{D}^x_{d^*} g^{ce^*} \bar{D}^x_{e^*}
             \label{derivativeV0},
    \end{eqnarray}
  where  we have used (\ref{cdf0}),(\ref{cdfn})  in the last equality.
  Thus, the first vacuum condition is
    \begin{equation}
      \langle \partial_c V \rangle
       =     \langle
             \partial_c 
             \left(
             e^{\CK} g_{ab^*} \CD^a \CD^b
             \right)
             \rangle
           + \langle \frac{e^{\mathcal{K}}}{(b^0)^2} i C_{acd} g^{ab^*} \bar{D}^x_{b*} g^{de^*} \bar{D}^x_{e*} \rangle
       =     0,
             \label{VC1}
    \end{equation}
  The second vacuum condition is to be with respect to the hypermultiplet scalar $ b^u $.
  As the potential contains only $ b^0 $, the condition reads
    \BEA
      \langle \frac{\partial V}{\partial b^0} \rangle
      &=&  - \frac{e^{\mathcal{K}} }{(b^0)^3}
             \langle
             2 g^{ab^*} D^x_a \bar{D}^x_{b^*}
           - (\mathcal{E}^x + \mathcal{M}^x \CF_n)(\mathcal{E}^x + \mathcal{M}^x \bar{\CF}_n)
             \rangle
       =     0. 
             \label{VC2}
    \EEA

  As we search for the vacua  with unbroken $SU(N)$ gauge symmetry in this paper, 
  we  will work on the condition
  $ \langle z^a \rangle = \delta^{an} \lambda $.
  Then $ \langle \CD^a \rangle = \langle \frac{i}{\sqrt{2}} f^a_{~bc} \bar{z}^{b^*} z^c \rangle = 0 $  holds and
  $ \langle \partial_c (e^{\CK} g_{ab^*} \CD^a \CD^b) \rangle = 0 $.
  For concreteness, we assume a form of the gauge invariant function $\CF(z)$ 
  as the one which parallels that of \cite{FIS3}:
    \begin{eqnarray}
      \CF(z)
      &=& - \frac{i C}{2} (z^n)^2 + \mathcal{G}(z),
            \\
      \mathcal{G}(z)
      &=&   \sum_{l=0}^k \frac{C_l}{l!} tr (z^a t_a)^l,
    \end{eqnarray}
  where $ C \in \mathbb{R} $ and  $C_l$ are constant.
  We will see that $ C $ must be nonvanishing in order for
  the inverse of the K\"ahler metric to exist.
  
  Let us compute the expectation value of the derivative of $ \CF $
    \BEA
      \langle \CF_a \rangle
      &=&    \delta_{an} \langle \CF_n \rangle,
             \nonumber \\
      \langle \CF_{na} \rangle
      &=&    \delta_{an} \langle \CF_{nn} \rangle,
             \nonumber \\
      \langle \CF_{\hat{a} \hat{b}} \rangle
      &=&    \delta_{\hat{a} \hat{b}} \langle \CF_{nn} - i C \rangle,
             \nonumber \\
      \langle \CF_{nab} \rangle
     &=&     \delta_{ab} \langle \CF_{nnn} \rangle,
             \label{fderivative}
    \EEA
  where the explicit form of $ \langle \CF_n \rangle $ and
   that of $ \langle \CF_{nn} \rangle $ are respectively
    \BEA
      \langle \CF_n \rangle
      &=&    \sum_{l} \frac{C_l}{(l-1)!} \lambda^{l-1} + i C \lambda,
             \nonumber \\
      \langle \CF_{nn} \rangle
      &=&    \sum_{l} \frac{C_l}{(l-2)!} \lambda^{l-2}.
    \EEA
  It is easy to compute $ \partial_a \mathcal{K} $,
    \BEA
      \langle \partial_a \mathcal{K} \rangle
      &=&  - \frac{i \langle e^{\mathcal{K}} \rangle}{2} 
             \langle 
             \CF_a - \bar{\CF}_{a} - (\lambda - \bar{\lambda}) \CF_{an} 
             \rangle
             \nonumber \\
      &=&    \delta_{an} \langle \partial_n \mathcal{K} \rangle.
             \label{partialK}
    \EEA
  The K\"ahler metric $ g_{ab} $ is
    \BEA
      \langle g_{ab^*} \rangle
      &=&    \left(
             \begin{array}{ccccc}
             \langle g_{11^*} \rangle &&&&
             \\
             & \langle g_{11*} \rangle &&& 0
             \\
             &&\ddots&
             \\
             &&&\ddots&
             \\
             0 &&&& \langle g_{nn} \rangle
             \end{array}
             \right),
    \EEA
  with
    \BEA
      \langle g_{11*} \rangle
      &=&  - \frac{i \langle e^{\mathcal{K}} \rangle}{2} \langle \CF_{nn} - \bar{\CF}_{nn} - 2iC \rangle,
             \nonumber \\
      \langle g_{nn} \rangle
      &=&    |\langle \partial_n \mathcal{K} \rangle|^2
           - \frac{i \langle e^{\mathcal{K}} \rangle}{2} \langle \CF_{nn} - \bar{\CF}_{nn} \rangle.
             \label{gVE1}
    \EEA
  Note that the diagonal components except $ \langle g_{nn} \rangle $ take the same value.
  By substituting  the above values, $ \langle D^x_a \rangle $ and $ \langle C_{abc} \rangle $ 
  take the following expression:
    \BEA
      \langle D^x_a \rangle
      &=&    \delta_{an}
             \frac{1}{\sqrt{2}} 
             \langle
             \mathcal{E}^x \partial_n \mathcal{K} + \mathcal{M}^x (\CF_{nn} + \partial_n \mathcal{K} \CF_n)
             \rangle,
             \nonumber \\
      &=&    \delta_{an} \langle D^x_n \rangle
             \nonumber \\
      \langle C_{abc} \rangle
      &=&    \frac{\langle e^{\mathcal{K}} \rangle}{2} \langle \CF_{abc} \rangle.
             \label{Cabc}
    \EEA

  Now we are ready to analyze (\ref{VC1}) and (\ref{VC2}).
  Substituting (\ref{fderivative})-(\ref{Cabc}) into (\ref{VC1}), we obtain
    \BEA
      \langle \frac{i e^{2 \mathcal{K}}}{2 (b^0)^2} \CF_{nnn} g^{n n^*} \bar{D}^x_{n^*} g^{n n^*} \bar{D}^x_{n^*} \rangle
       =     0.
    \EEA
  The points $ \langle \CF_{nnn} \rangle = 0 $ are unstable vacua 
  because $ \langle \partial_{a} \partial_{b^*} V \rangle = 0 $.
  The points which satisfy $ \langle g^{n n^*} \rangle = 0 $ 
  and $ \langle \partial_n \mathcal{K} \rangle = 0 $ are not stable.
  The vacuum condition reduces to
    \begin{equation}
      \langle \bar{D}^x_{n^*} \bar{D}^x_{n^*} \rangle
       =     0,
             \label{VC3}
    \end{equation}
  which implies
    \begin{equation}
      \LLangle \CF_n + \frac{\CF_{nn}}{\partial_n \mathcal{K}} \RRangle
       =   - \left(
             \frac{g_2}{g_3} \pm i \frac{g_1}{g_3}
             \right).
             \label{VC4}
    \end{equation}
  where we use $ \llangle \cdots \rrangle $ for the vacuum expectation value
  of $ \cdots$ 
  which are determined as the solutions to (\ref{VC3}).
  We have also assumed $ g_3 \neq 0 $.
  Note that if $ g_3 = 0 $ (\ref{VC3}) leads to $ g_1 = g_2 = 0 $ and the supersymmetry is unbroken.
  
  The second condition (\ref{VC2}) reads
    \BE
      \left|
      g_1 \mp i g_3 \LLangle \frac{\CF_{nn}}{\partial_n \mathcal{K}} \RRangle
      \right|^2 + g_1^2 - \llangle g^{nn^*} |\partial_n \mathcal{K}|^2 \rrangle 2 g_1^2
       =    0.
            \label{VC5}
    \EE
  When $ \llangle \CF_{nn} \rrangle = 0 $, 
  (\ref{gVE1}) imply $ \llangle g^{nn^*} |\partial_n \mathcal{K}|^2 \rrangle = 1 $.
  Thus the above equality is satisfied.
  On the other hand, when $ \llangle \CF_{nn} \rrangle \neq  0 $, (\ref{VC5}) leads to $ g_3 =0 $.
  This is proven in the appendix.
  $ g_3 =0 $ conflicts with the assumption.
  The second vacuum condition thus reduces to
    \BE
      \llangle \CF_{nn} \rrangle = 0.
           \label{VC6}
    \EE
  In the rigid theory \cite{FIS12, FIS3} with no hypermultiplet, there is no counterpart to this equation.
  In fact, we will see shortly that $ \mathcal{N} = 2 $ local supersymmetry is not broken partially 
  without invoking the second vacuum condition.
   We conclude from  (\ref{VC4})  (\ref{VC6})
    \BE
      \llangle \CF_n \rrangle
       =   - \left(
             \frac{g_2}{g_3} \pm i \frac{g_1}{g_3}
             \right).
             \label{VC7}
    \EE
  In what follows, we take the $+$ sign.
  So $ \partial_a \mathcal{K} $ and $ g_{ab^*} $ take the following expression:
   \BEA
      \llangle \partial_a \mathcal{K} \rrangle
      &=&  - \delta_{an} \llangle e^{\mathcal{K}} \rrangle \frac{g_1}{g_3},
             \nonumber \\
      \llangle g_{11^*} \rrangle
      &=&  - \llangle e^K \rrangle C,
             \nonumber \\
      \llangle g_{nn^*} \rrangle
      &=&    |\llangle \partial_n \mathcal{K} \rrangle|^2
       =     \llangle e^{2 \mathcal{K}} \rrangle 
             \left(
             \frac{g_1}{g_3}
             \right)^2.
    \EEA
  Note that $ C \neq 0 $ is necessary  for the K\"ahler metric
  to be invertible.
  
  Let us see if extended supersymmetries are spontaneously broken or not
  by considering the vacuum expectation values of the
  mass matrices (\ref{S})-(\ref{N}).  They are
    \begin{eqnarray}
      \llangle S_{AB} \rrangle
      &=&  - \LLangle
             \frac{i e^{\mathcal{K}/2}}{2 \sqrt{2} b^0} g_1 
             \RRangle
             \left(
             \begin{array}{cc}
             1 & 1 
             \\
             1 & 1
             \end{array}
             \right),
             \label{Svev}
             \nonumber \\
      \llangle W^{aAB} \rrangle
      &=&    \delta^{an}
             \LLangle
             \frac{i e^{\mathcal{K}/2}}{\sqrt{2} b^0} (\partial_n \mathcal{K})^{-1} g_1\RRangle
             \left(
             \begin{array}{cc}
             1 & 1
             \\
             1 & 1
             \end{array}
             \right),
             \label{Wvev}
             \nonumber \\
      \llangle N^{A}_{\alpha} \rrangle
      &=&    \LLangle
             \frac{i e^{\mathcal{K}/2}}{\sqrt{2} b^0} g_1 
             \RRangle
             \left(
             \begin{array}{cc}
             1 & 1
             \\
             -1 & -1
             \end{array}
             \right).
             \label{Nvev}
    \end{eqnarray}
  Notice that these matrices have zero eigenvalues.
  Introducing $ \phi_{\pm} = \frac{1}{\sqrt{2}} (\phi_1 \pm \phi_2) $
  where $ \phi \in {\psi, \zeta, \lambda} $, we obtain  the vevs of (\ref{strns1}), (\ref{strns2}), (\ref{strns3}). 
    \BEA
      \llangle \delta \psi_{+ \mu} \rrangle
      &=&    \LLangle
             \frac{i e^{\mathcal{K}/2}}{2 b^0} g_1 
             \RRangle
             \gamma_\mu (\epsilon_1 + \epsilon_2)
             \nonumber \\
      \llangle \delta \lambda^{a +} \rrangle
      &=&    \delta^{an}
             \LLangle
             \frac{i e^{\mathcal{K}/2}}{ b^0} g_1 (\partial_n \mathcal{K})^{-1}
             \RRangle
             (\epsilon_1 + \epsilon_2)
             \nonumber \\
      \llangle \delta \zeta_{-} \rrangle
      &=&    \LLangle
             \frac{i e^{\mathcal{K}/2}}{b^0} g_1 
             \RRangle
             (\epsilon_1 + \epsilon_2)
             \nonumber \\
      \llangle \delta \psi_{- \mu} \rrangle
      &=&    \llangle \lambda^{a -} \rrangle
       =     \llangle \zeta_{+} \rrangle
       =     0.
    \EEA
  Let us further introduce
    \BEA
      \chi_\bullet
      &\equiv&    \llangle \partial_n \mathcal{K} \rrangle \lambda^{n +} + 2 \zeta_{-},
             \nonumber \\
      \eta_\bullet
      &\equiv&  - \llangle \partial_n \mathcal{K} \rrangle \lambda^{n +} + \zeta_{-},
    \EEA
  whose supersymmetry transformations are
    \BEA
      \llangle \delta \chi_\bullet \rrangle
      &=&    \LLangle
             \frac{3 i e^{\mathcal{K}/2}}{b^0} g_1 
             \RRangle
             (\epsilon_1 + \epsilon_2),
             \nonumber \\
      \llangle \delta \eta_\bullet \rrangle
      &=&    0,
             \label{SUSYtr}
    \EEA
  where the upper and lower position of dot represent left and right chirality respectively.
  As we see in the next section, 
  gravitino $ \psi_{-} $, hyperino $ \zeta_{+} $, gaugino $ \lambda_{-}^a $ and $ \chi $ are massless fermions  while
  gravitino $ \psi_{+} $, gaugino $ \lambda_{+}^{\hat{a}} $ and $ \eta $ are massive physical fermions.
  $ \mathcal{N} = 2 $ local supersymmetry is spontaneously broken to $ \mathcal{N} = 1 $ 
  and $ \chi $ is the Nambu-Goldstone fermion.
  We will confirm this  in the next section.
    
\section{Mass Spectrum}

\subsection{Fermion Mass}
  Let us consider the fermion mass spectrum.
  Substituting (\ref{Nvev}) into $ \CL_{\rm{Yukawa}} $,
  we obtain
    \BEA
      \CL_{\rm{Yukawa}}
      &=&  - i \LLangle \frac{\sqrt{2}e^{\mathcal{K}/2}}{b^0} g_1 \RRangle
             \left(
             \bar{\psi}^{+}_\mu \gamma^{\mu \nu} \psi^{+}_\nu
           - i \bar{\chi}^\bullet \gamma_\mu \psi^\mu_{+}
           + \frac{1}{3} \bar{\chi}_\bullet \chi_\bullet
           - \frac{1}{3} \bar{\eta}_\bullet \eta_\bullet
             \right)
             \nonumber \\
      & &  + \frac{1}{2 \sqrt{2}} \LLangle \frac{e^{\mathcal{K}/2}}{b^0} g_3 \CF_{aan} \RRangle 
             \bar{\lambda}^{a -} \lambda^{a -}
           + \ldots + h.c.,
    \EEA
  The Nambu-Goldstone fermion $ \chi $ coupling to the gravitino $\psi^{+} $
  can be removed from the action by redefining the gravitino:
    \BE
      \psi^+_\mu
       \rightarrow \psi^+_\mu + \frac{i}{6} \gamma_{\mu} \chi_\bullet.
    \EE
  We obtain
    \BE
      \CL_{\rm{Yukawa}}
       =   - i  \LLangle \frac{\sqrt{2}e^{\mathcal{K}/2}}{b^0} g_1 \RRangle
             \left(
             \bar{\psi}^{+}_\mu \gamma^{\mu \nu} \psi^{+}_\nu
           - \frac{1}{3} \bar{\eta}_\bullet \eta_\bullet
             \right)
           + \frac{1}{2 \sqrt{2}} \sum_{a=1}^n \LLangle \frac{e^{\mathcal{K}/2}}{b^0} g_3 \CF_{aan} \RRangle 
             \bar{\lambda}^{a -} \lambda^{a -}
           + h.c.~.
    \EE
  The gravitino $ \psi_+ $ has acquired a mass by the
  super-Higgs mechanism.

  The kinetic terms of the massive fermions are
    \BE
      \CL_{\rm{kin}}^{(f)}
       =     \frac{\epsilon^{\mu \nu \lambda \sigma}}{\sqrt{-g}}
             \bar{\psi}^A_\mu \gamma_\nu \partial_\lambda \psi_{A \sigma}
           - \frac{i}{3} \bar{\eta}^\bullet \gamma_\mu \partial^\mu \eta_\bullet
           - i \sum_{a=1}^n \llangle g_{aa^*} \rrangle \bar{\lambda}^{a -} \gamma_{\mu} \partial^{\mu} \lambda^{a^*}_- 
           + \ldots
           + h.c.
    \EE
  We obtain the mass of the fermions from the equations of motion.
  The gravitino mass $ m $ and the mass of the gauginos $ m_{a} $ are
  respectively  
  \BEA
      m
      &=&    \left|
             \LLangle 
             \frac{\sqrt{2}e^{\mathcal{K}/2}}{b^0} g_1 
             \RRangle 
             \right|,
             \label{m}
             \nonumber \\
      m_a
      &=&    \left| 
             \LLangle 
             \frac{e^{\mathcal{K}/2}}{\sqrt{2 }b^0} g_3 \CF_{aan} g^{aa^*} 
             \RRangle 
             \right|.
             \label{ma}
    \EEA
  Notice that the mass of the physical fermion $ \eta_\bullet $ is the
  same as the gravitino, that is, $ m $.
  $ \psi_+ $ and $ \eta_\bullet $ will form a $ \mathcal{N} =1 $ massive multiplet of spin (3/2,1,1,1/2),
  while $ \lambda^{a -} $, together with the scalar fields, 
  will form $ \mathcal{N} =1 $ massive chiral multiplet.
  This mass spectrum is analogous to the rigid counterpart \cite{FIS3}.

\subsection{Boson Mass}
  Let us compute the masses of the scalar fields
  by introducing the shifted fields  $\tilde{z}^a =  z^a - \llangle z^a \rrangle$.
  The second derivatives can be easily evaluated
    \BEA
      \llangle \partial_a \partial_{b*} V \rrangle
      &=&   \LLangle
            \frac{2 i e^{\mathcal{K}}}{(b^0)^2} 
            C_{acd} g^{ce^*} (\partial_{b^*} \bar{D}^x_{e^*}) g^{df^*} \bar{D}^x_{f^*}
            \RRangle
            \nonumber \\
      &=&   \delta_{ab}
            \LLangle
            \frac{e^{\mathcal{K}}}{2 (b^0)^2} |g_3 \CF_{aan}|^2 g^{aa^*}
            \RRangle,
            \nonumber \\
      \llangle \partial_a \partial_b V \rrangle
      &=&    0.
    \EEA
  Thus, the kinetic terms and the mass terms are
    \BE
            \sum_a 
            \left(
            \llangle g_{aa^*} \rrangle \partial_\mu \tilde{z}^a \partial^\mu \bar{\tilde{z}}^{a^*}
          - \LLangle
            \frac{e^{\mathcal{K}}}{2 (b^0)^2} |g_3 \CF_{aan}|^2 g^{aa^*}
            \RRangle
            \tilde{z}^a \bar{\tilde{z}}^{a^*}
            \right).
    \EE
  The mass of $ \tilde{z}^a $ is the same as (\ref{ma}), namely, the mass of $ \lambda^{a -} $.
  They form $ N^2 $ massive chiral multiplets as we have anticipated,

  The gauge boson masses appear in the kinetic terms of the hypermultiplet scalars
    \BEA
      h_{uv} \nabla_\mu b^u \nabla^\mu b^v
      &=&    \frac{1}{2 (b^0)^2} 
             (g_1^2 A^0_\mu A^{0 \mu} + g_3^2 A'^n_\mu A'^{n \mu})
           + \ldots,
             \label{Amass}
    \EEA
  where
    \BE
      A^{\prime n}_\mu
       =     A^n_\mu 
           + \left( \frac{g_2}{g_3} \right) A^0_\mu
             \label{A'}.
    \EE
  The kinetic terms of the gauge bosons are 
  $ \frac{1}{4} (\Im \mathcal{N})_{\Lambda \Sigma} F^\Lambda_{\mu \nu} F^{\Sigma \mu \nu} $, 
  and  we  compute the generalized coupling matrix $ \mathcal{N} $ (\ref{CouplingMatrix}) on the vacuum:
    \BE
      \llangle \mathcal{N}_{\Lambda \Sigma} \rrangle
       =    \begin{pmatrix}
            \llangle \mathcal{N}_{00} \rrangle & 0 & \cdots & \cdots & 0 & \llangle \mathcal{N}_{0n} \rrangle
            \\
            0 & \llangle \mathcal{G}_{11} \rrangle & 0 & \cdots & 0 & 0
            \\
            \vdots & 0 & \llangle \mathcal{G}_{22} \rrangle & & \vdots & \vdots & 
            \\
            \vdots & \vdots & & \ddots & 0 & \vdots
            \\
            0 & 0 & \cdots & 0 & \llangle \mathcal{G}_{n-1,n-1} \rrangle & 0
            \\
            \llangle \mathcal{N}_{n0} \rrangle & 0 & \cdots & \cdots & 0 & \llangle \mathcal{N}_{nn} \rrangle
            \\
            \end{pmatrix},
    \EE
  with
    \BEA
      \Im \llangle \mathcal{N}_{00} \rrangle
      &=&    \LLangle
             \frac{e^{-\mathcal{K}}}{2} 
             \RRangle \frac{g_1^2 + g_3^2}{g_1^2},
             \nonumber \\
      \Im \llangle \mathcal{N}_{0n} \rrangle
      &=&    \Im \llangle \mathcal{N}_{n0} \rrangle
       =     \LLangle
             \frac{e^{-\mathcal{K}}}{2} 
             \RRangle \frac{g_2 g_3}{g_1^2},
             \nonumber \\
      \Im \llangle \mathcal{N}_{nn} \rrangle
      &=&    \LLangle
             \frac{e^{-\mathcal{K}}}{2} 
             \RRangle 
             \left(
             \frac{g_3}{g_1}
             \right)^2.
    \EEA
  Therefore the gauge boson kinetic terms are
    \BEA
      \frac{1}{4} \Im \llangle \mathcal{N}_{\Lambda \Sigma} \rrangle F^\Lambda_{\mu \nu} F^{\Sigma \mu \nu}
      &=&  - \LLangle
             \frac{e^{-\mathcal{K}}}{8} 
             \RRangle
             F^0_{\mu \nu} F^{0 \mu \nu}
           - \LLangle
             \frac{e^{-\mathcal{K}}}{8} 
             \RRangle
             \left(\frac{g_3}{g_1}\right)^2 F'^n_{\mu \nu} F'^{n \mu \nu}  
             \nonumber \\
      & &  + \frac{1}{4} \sum_{\hat{a}} \Im \llangle \mathcal{G}_{\hat{a} \hat{a}} \rrangle F^{\hat{a}}_{\mu \nu} F^{\hat{a} \mu \nu}
             \label{Akin},
    \EEA
  where we have defined $ F'^{n}_{ \mu \nu} = \partial_\mu A'^{n}_\nu 
- \partial_\nu A'^{n}_\mu $.
  We can read off the masses of gauge boson $ A^0_\mu $ and $ A'^n_\mu $ from (\ref{Amass}) and (\ref{Akin}).
  Both of them agree with (\ref{m}).
  
  We summarize the mass spectrum of our model in the table below:

    \begin{center}
      \begin{tabular}{|c|c|c|}
        \hline
        $ \CN = 1 $ multiplet & field & mass \\ \hline \hline
        gravity multiplet & $ e^a_{\mu} $, $ \psi^{-}_{\mu} $ & 0 \\ \hline
        spin-3/2 multiplet & $ \psi^{+}_{\mu} $, $ A^0_\mu $, $ A'^n_\mu $, $ \eta_\bullet $ & $ m $ \\ \hline
        $ SU(N) $ vector multiplet & $ A^{\hat{a}}_\mu $, $ \lambda^{\hat{a} +} $ & 0 \\ \hline
        $ SU(N) $ adjoint chiral multiplet & $ \lambda^{\hat{a} -} $, $ z^{\hat{a}} $ & $ m^{\hat{a}} $ \\ \hline
        chiral multiplet & $ \lambda^{n -} $, $ z^{n} $ & $ m^n $ \\ \hline
        chiral multiplet & $ \zeta_+ $, $ b^0,b^1 $ & 0 \\
        \hline
      \end{tabular}
    \end{center}
  The $\CN=1$ gravity multiplet consists of the vierbein and the gravitino $ \psi^-_\mu $ while 
  the massive gravitino $ \psi^+_\mu $, $ U(1) $ gauge boson $ A'^n_\mu $, the graviphoton $ A^0_\mu $ 
  and the fermion $ \eta_\bullet $ form a massive spin-3/2 multiplet.
  The $\CN=2$ $U(N)$ vector multiplet is  divided into a $\CN=1$ vector multiplet and a chiral multiplet.
  The $ \mathcal{N} = 1$  $SU(N)$ vector multiplet consists of massless gauge bosons $ A^{\hat{a}}_\mu $
  and gauginos $ \lambda^{\hat{a} +} $.
  On the other hand, the gaugino $ \lambda^{\hat{a} -} $ and the scalar field $ z^{\hat{a}} $ form chiral multiplets 
  which belong to the $ SU(N) $ adjoint representation.
  The hyperino $ \zeta_+ $ and the scalars $ b^0,b^1 $ form an $ \mathcal{N} = 1 $ chiral multiplet.
  
  Note that the $U(1)_{graviphoton} \times U(N)$ gauge symmetry is broken to $SU(N)$
  and the vacuum lies in the Higgs phase.
  
\section{$\CN = 1$ Lagrangian}
  In the last section, we have considered the lowest order terms 
  with respect to the fermion fields and the shifted scalar fields $ \tilde{z}^a $ in $\CL_{\rm{Yukawa}}$ and $V$.
  We will now reexpress the remaining terms in $\CL_{\rm{Yukawa}}$ and $V$ by $\tilde{z}^a $ as well. 
  In \cite{Andri1, Andri2}, the reduction procedure from $\CN=2$ to $\CN=1$ has been completed 
  and $\CL_{\rm{Yukawa}}$ and $V$ have been given in terms of a superpotential.
  Here, we find that the resulting $\CN = 1$ Lagrangian on the vacuum is written by the superpotentials
  which are related each other by eq. (\ref{superpotential}).
  
  The holomorphic function $\CF(z)$ is expanded in the shifted fields $\tilde{z}^a$ as,
    \beq
      \CF (z)
      =   \CF (\llangle z \rrangle + \tilde{z})        
      =   \llangle \CF \rrangle + \tilde{\CF},
    \eeq
  where 
    \BE
      \tilde{\CF}
       =    \llangle \CF_a \rrangle \tilde{z}^a + \frac{1}{2!} \llangle \CF_{ab} \rrangle \tilde{z}^a \tilde{z}^b
          + \frac{1}{3!} \llangle \CF_{abc} \rrangle \tilde{z}^a  \tilde{z}^b \tilde{z}^c
          + \ldots.
    \EE
  Similarly, $ \CF_a $ and $ \CF_{ab} $ are
    \beq
      \CF_a
       =    \llangle \CF_a \rrangle + \llangle \CF_{ab} \rrangle \tilde{z}^b
          + \frac{1}{2!} \llangle \CF_{abc} \rrangle \tilde{z}^b \tilde{z}^c
          + \ldots.
       =    \tilde{\CF_a},
            \\
      \CF_{ab}
       =    \llangle \CF_{ab} \rrangle + \llangle \CF_{abc} \rrangle \tilde{z}^c
          + \ldots.
       =    \tilde{\CF_{ab}}.
    \eeq
  The derivatives  are taken with respect to  $ \tilde{z}^a $ in $ \tilde{\CF}_a $ and $ \tilde{\CF}_{ab} $.
  The K\"ahler potential and its derivatives are
    \BEA
      \CK
      &=& - \log i 
            \left[
            \llangle \CF - \bar{\CF} \rrangle + \tilde{\CF} - \bar{\tilde{\CF}} 
          - \frac{1}{2} ( \llangle z^a - \bar{z}^a \rrangle + z^a - \bar{z}^a ) (\tilde{\CF_a} + \bar{\tilde{\CF_a}})
            \right],
            \\
      \partial_a \CK
      &=& - \frac{i}{2 \CK_0} (\tilde{\CF_a} - \bar{\tilde{\CF_a}} 
          - ( \llangle z^a - \bar{z}^a \rrangle + z^a - \bar{z}^a ) \tilde{\CF}_{ab}) 
            =   \tilde{\partial_a} \CK,
            \\
      g_{ab^*}
      &=&   \tilde{\partial_a} \CK \tilde{\partial}_{b^*} \CK 
          - \frac{i}{2 \CK_0} ( \tilde{\CF}_{ab} - \bar{\tilde{\CF}}_{ab})
           =   \tilde{g}_{ab^*},
    \EEA
  where $ \tilde{\partial_a} = \partial/\partial \tilde{z}^a $.

  Let us now define the `two' superpotentials by
    \BEA
      \CW (\tilde{z}, \bar{\tilde{z}})
      &\equiv&  
            e^{\CK/2} W(\tilde{z})
       \equiv
            2(S_{11} - S_{12})
       =    \frac{e^{\CK/2}}{\sqrt{2} b^0} g_3 (\tilde{\CF_n} - \llangle \CF_n \rrangle),
            \\
      \CS (\tilde{z}, \bar{\tilde{z}})
      &\equiv&  
            e^{\CK/2} S(\tilde{z})
       \equiv
            2(S_{11} + S_{12})
       =    \frac{e^{\CK/2}}{\sqrt{2} b^0} ( 2 g_2 + g_3 (\tilde{\CF_n} + \llangle \CF_n \rrangle) ),
    \EEA
  where $ S_{AB} $ is the gravitino mass matrix.
  Note that $ \CW $ and $ \CS $ are related as follows:
    \BE
      \CW
       =    \CS + i \frac{\sqrt{2} g_1 e^{\CK/2}}{b^0}.
            \label{superpotential}
    \EE
  Thus, they are not independent quantities.
  In the following, however, we will write down the resulting Lagrangian, using both $\CW$ and $\CS$.
  These quantities appear in the gravitino mass terms as
    \BEA
      2 S_{AB} \bar{\psi}^A_\mu \gamma^{\mu \nu} \psi^B_\nu 
      &=&   \CW \bar{\psi}^{-}_{\mu} \gamma^{\mu \nu} \psi^{-}_{\nu}
          + \CS \bar{\psi}^{+}_{\mu} \gamma^{\mu \nu} \psi^{+}_{\nu}.
    \EEA
  The covariant derivative of $ \CW $ and that of $ \CS $ are respectively
    \BEA
      \tilde{\nabla}_a\CW
      &=&   \frac{e^{\CK/2}}{\sqrt{2} b^0} g_3 
            ( \tilde{\CF}_{na} + \tilde{\partial_a} \CK \tilde{\CF_n} - \tilde{\partial_a} \CK \llangle \CF_n \rrangle )
            \nonumber \\
      &=&   g_{ab^*} (\bar{W}^{b^*}_{2;11} - \bar{W}^{b^*}_{2;12}),
            \\
      \tilde{\nabla}_a \CS
      &=&   \frac{e^{\CK/2}}{\sqrt{2} b^0} 
            (2 g_2 \tilde{\partial_a} \CK + g_3 
            ( \tilde{\CF_{na}} + \tilde{\partial_a} \CK \tilde{\CF_n} + \tilde{\partial_a} \CK \llangle \CF_n \rrangle ))
            \nonumber \\
      &=&   g_{ab^*} (\bar{W}^{b^*}_{2;11} + \bar{W}^{b^*}_{2;12}),
            \\ 
      \tilde{\nabla}_a \tilde{\nabla}_b \CW
      &=&   \sqrt{2} (\mathcal{M}_{2;a1 \vert b1} - \mathcal{M}_{2;a1 \vert b2}),
            \\
      \tilde{\nabla}_a \tilde{\nabla}_b \CS
      &=&   \sqrt{2} (\mathcal{M}_{2;a1 \vert b1} + \mathcal{M}_{2;a1 \vert b2}),
    \EEA
  where $ \bar{W}^{b^*}_{AB} = (W^{bAB})^* $.
  These are used to evaluate the second term and the last term of $ \CL_{\rm{Yukawa}} $:
    \BEA
      i g_{ab^*} W^{aAB} \bar{\lambda}^{b^*}_A \gamma_\mu \psi^\mu_B
      &=&   e^{\CK/2} \tilde{g}_{ab^*} \CD^a 
            ( \bar{\lambda}^{b^*}_{-} \gamma_\mu \psi^\mu_{+} - \bar{\lambda}^{b^*}_{+} \gamma_\mu \psi^\mu_{-} )
            \nonumber \\
      & & ~~~+ ~i \bar{\tilde{\nabla}}_{a^*} \bar{\CW} \bar{\lambda}^{b^*}_{-} \gamma_\mu \psi^\mu_{-}
          + i \bar{\tilde{\nabla}}_{a^*} \bar{\CS} \bar{\lambda}^{b^*}_{+} \gamma_\mu \psi^\mu_{+},
            \\
      \mathcal{M}_{aA \vert bB} \bar{\lambda}^{aA} \lambda^{bB}
      &=&   \mathcal{M}_{1;a1 \vert b2} (\bar{\lambda}^{a-} \lambda^{b+} - \bar{\lambda}^{a+} \lambda^{b-})
            \nonumber \\
      & & ~~~+ ~\frac{1}{\sqrt{2}} \tilde{\nabla}_a \tilde{\nabla}_b \CW \bar{\lambda}^{a-} \lambda^{b-}
          + \frac{1}{\sqrt{2}} \tilde{\nabla}_a \tilde{\nabla}_b \CS \bar{\lambda}^{a+} \lambda^{b+}.
    \EEA
  We  now manage to reexpress  $ \CL_{\rm{Yukawa}} $
  by the shifted scalar fields, the superpotentials and their covariant derivatives:
    \BEA
      \CL_{\rm{Yukawa}}
      &=&   \CW \bar{\psi}^{-}_{\mu} \gamma^{\mu \nu} \psi^{-}_{\nu}
          + i (\bar{\tilde{\nabla}}_{a^*} \bar{\CW} \bar{\lambda}^{a^*}_{-} 
          - 2 \bar{\CW} \bar{\zeta}^{+}) \gamma_\mu \psi^\mu_{-}
          - e^{\CK/2} \tilde{g}_{ab^*} \CD^a \bar{\lambda}^{b^*}_{+} \gamma_\mu \psi^\mu_{-}
            \nonumber \\
      & & + \CS \bar{\psi}^{+}_{\mu} \gamma^{\mu \nu} \psi^{+}_{\nu}
          + i (\bar{\tilde{\nabla}}_{a^*} \bar{\CS} \bar{\lambda}^{a^*}_{+} 
          + 2 \bar{\CS} \bar{\zeta}^{-}) \gamma_\mu \psi^\mu_{+}
          + e^{\CK/2} \tilde{g}_{ab^*} \CD^a \bar{\lambda}^{b^*}_{-} \gamma_\mu \psi^\mu_{+}
            \nonumber \\
      & & + \CW \bar{\zeta}_+ \zeta_+ 
          + \CS \bar{\zeta}_- \zeta_-
          - 2 \tilde{\nabla}_a \CW \bar{\zeta}_+ \lambda^{a-}
          + 2 \tilde{\nabla}_a \CS \bar{\zeta}_- \lambda^{a+}
            \label{finalmass} \\
      & & + \mathcal{M}_{1;a1 \vert b2} (\bar{\lambda}^{a-} \lambda^{b+} - \bar{\lambda}^{a+} \lambda^{b-})
          + \frac{1}{\sqrt{2}} \tilde{\nabla}_a \tilde{\nabla}_b \CW \bar{\lambda}^{a-} \lambda^{b-}
          + \frac{1}{\sqrt{2}} \tilde{\nabla}_a \tilde{\nabla}_b \CS \bar{\lambda}^{a+} \lambda^{b+}
          + h.c.~.
            \nonumber
    \EEA

 Let us turn to the scalar potential $ V $, which we rewrite in terms of
 the mass matrices as
    \BE
      V
       =  - 12 \bar{S}^{1A} S_{A1} + \tilde{g}_{ab^*} \bar{W}^{b^*}_{1A} W^{a1A}
          + 2 \bar{N}_1^{\alpha} N_{\alpha}^1.
            \label{V}
    \EE
  This equation is also obtained from the supergravity Ward identities in the reference \cite{Cecotti}.
  Note that $ \bar{S}^{AB} = (S_{AB})^* $ and $ \bar{N}_A^\alpha = (N_\alpha^A)^* $.
  The first term is
    \BEA
      -12 \bar{S}^{1A} S_{A1}
      &=& - 6 ((S_{11} - S_{12})(\bar{S}^{11} - \bar{S}^{12}) + (S_{11} + S_{12})(\bar{S}^{11} + \bar{S}^{12}))
            \nonumber \\
      &=& - \frac{3}{2} (|\CW|^2 + |\CS|^2),
            \label{SS}
    \EEA
  and the second term is
    \BEA
      \tilde{g}_{ab^*} \bar{W}^{b^*}_{1A} W^{a1A}
      &=&   \tilde{g}_{ab^*} 
            \left(
            \bar{W}^{b^*}_{1;12} W^{a12}_1 + \bar{W}^{b^*}_{2;11} W^{a11}_2 + \bar{W}^{b^*}_{2;12} W^{a12}_2
            \right)
            \nonumber \\
      &=&   e^{\CK} \tilde{g}_{ab^*} \mathcal{D}^a \mathcal{D}^b
          + \frac{1}{2} \tilde{g}^{ab^*} \tilde{\nabla}_a \CW \bar{\tilde{\nabla}}_a \bar{\CW}
          + \frac{1}{2} \tilde{g}^{ab^*} \tilde{\nabla}_a \CS \bar{\tilde{\nabla}}_a \bar{\CS}.
    \EEA
  In the first equality, we have used (\ref{killing}).
  The last term is
    \BEA
      2 \bar{N}_1^{\alpha} N_{\alpha}^1
      &=&   |\CW|^2 + |\CS|^2
            \nonumber \\
      &=&   \frac{1}{2} h^{uv} \nabla_u \CW \nabla_v \bar{\CW}
          + \frac{1}{2} h^{uv} \nabla_u \CS \nabla_v \bar{\CS},
            \label{NN}
    \EEA
  where $ u, v = 0,1 $ and $ h_{uv} \equiv \delta_{uv}/2(b^0)^2 $.
  Note that $ \nabla_u \CW = \partial_u \CW $.
  Substituting (\ref{SS})-(\ref{NN}) into (\ref{V}), we obtain
    \BEA
      V
      &=&   e^{\CK/2} g_{ab^*} \mathcal{D}^a \mathcal{D}^b
          + \frac{1}{2} \tilde{g}^{ab^*} \tilde{\nabla}_a \CW \bar{\tilde{\nabla}}_a \bar{\CW}
          + \frac{1}{2} \tilde{g}^{ab^*} \tilde{\nabla}_a \CS \bar{\tilde{\nabla}}_a \bar{\CS}
            \nonumber \\
      & & - \frac{3}{2} |\CW|^2 - \frac{3}{2} |\CS|^2
          + \frac{1}{2} h^{uv} \nabla_u \CW \nabla_v \bar{\CW} 
          + \frac{1}{2} h^{uv} \nabla_u \CS \nabla_v \bar{\CS}.
            \label{finalpotential}
    \EEA
  This is the final form of the scalar potential.
  We can see that $\CL_{\rm{Yukawa}}$ and the scalar potential take essentially the same form as that of
  the usual $\CN=1$ supergravity models (such as \cite{WB} or \cite{Andri1, Andri2}).

  As is pointed out in \cite{Ferrara2}, if we force the gravity and the hypermultiplet to decouple, 
  the gravitino mass (\ref{m}) becomes zero.
  Thus, the gauge boson corresponding to the overall $U(1)$ and the graviphoton become massless in this limit.
  The Higgs phase of overall $U(1)_{graviphoton} \times U(1)$ approaches the Coulomb phase. 

\section{Acknowledgements}
  The authors thank Kazuhito Fujiwara, Makoto Sakaguchi and Takahiro Kubota for useful discussions.
  This work is supported in part by the Grant-in-Aid for Scientific Research (16540262) 
  from the Ministry of Education, Science and Culture, Japan.
  Support from the 21 century COE program ``Constitution of Wide-angle mathematical basis focused on knots'' 
  is gratefully appreciated.

\appendix

\section{}
  Here we prove the second vacuum condition (\ref{VC5}) leads to $\llangle \CF_{nn} \rrangle = 0$.
  As mentioned above, if $\llangle \CF_{nn} \rrangle = 0$, (\ref{VC5}) is automatically satisfied.
  Thus, let us consider the case $\llangle \CF_{nn} \rrangle \neq 0$.
  We write $\CF_{nn}$ as
    \BE
      \CF_{nn}
       =     F_1 + i F_2,
    \EE
  where $F_1, F_2 \in \mathbb{R}$.
  From (\ref{gVE1}), (\ref{partialK}) and (\ref{VC4}), by using $F_1$ and $F_2$, we obtain
    \BEA
      \llangle g_{nn^*} \rrangle
      &=&    \llangle |\partial_n \CK|^2 \rrangle
           + \frac{F_2}{\llangle \CK_0 \rrangle},
             \label{A5}
             \\
      \llangle \partial_n \CK + \partial_{n^*} \CK \rrangle
      &=&    \frac{i}{\llangle \CK_0 \rrangle}
             \left(
             \LLangle \frac{\CF_{nn}}{\partial_n \CK} - \frac{\bar{\CF}_{nn}}{\partial_{n^*} \CK} \RRangle
           \pm
             2i \frac{g_1}{g_3}
           + (\lambda - \bar{\lambda}) F_1
             \right),
             \label{A1}
             \\
      \llangle \partial_n \CK - \partial_{n^*} \CK \rrangle
      &=&    \frac{1}{\llangle \CK_0 \rrangle}
             (\lambda - \bar{\lambda}) F_2.
             \label{A2}
    \EEA
    
  The condition (\ref{VC5}) can be written as
    \BE
      0
       =     2 g_1^2
           - 2 g_1^2 \llangle g^{nn^*} |\partial_n \CK|^2 \rrangle
           \mp
             i Y g_1 g_3
           + g_3^2 \LLangle |\frac{\CF_{nn}}{\partial_n \CK}|^2 \RRangle,
             \label{A3}
    \EE
  where we have defined $Y$ as
    \BEA
      Y
      &\equiv&    
             \LLangle \frac{\CF_{nn}}{\partial_n \CK} - \frac{\bar{\CF}_{nn}}{\partial_{n^*} \CK} \RRangle
             \NN \\
      &=&    \frac{1}{\llangle |\partial_n \CK|^2 \rrangle}
             \left[
             F_1 \llangle \partial_n \CK - \partial_{n^*} \CK \rrangle
           - \frac{F_2}{\llangle \CK_0 \rrangle} 
             \left(
             Y \pm 2i \frac{g_1}{g_3} + (\lambda - \bar{\lambda}) F_1
             \right)
             \right].
    \EEA
  In the second equality, we have used (\ref{A1}).
  Using (\ref{A5}), we can solve the above equation for $Y$:
    \BE
      Y \llangle g_{nn^*} \rrangle
       =   \mp
             \frac{F_2}{\llangle \CK_0 \rrangle} 2i \frac{g_1}{g_3}.
             \label{A4}
    \EE
  Substituting (\ref{A4}) into (\ref{A3}), we get
    \BEA
      0
      &=&    2 g_1^2
           - 2 g_1^2 \llangle g^{nn^*} |\partial_n \CK|^2 \rrangle
           - 2 g_1^2 \LLangle g^{nn^*} \frac{F_2}{\llangle \CK_0 \rrangle} \RRangle
           + g_3^2 \LLangle |\frac{\CF_{nn}}{\partial_n \CK}|^2 \RRangle
             \NN \\
      &=&    g_3^2 \LLangle |\frac{\CF_{nn}}{\partial_n \CK}|^2 \RRangle,
    \EEA
  where we have used (\ref{A5}).
  Therefore, we conclude that when $\llangle \CF_{nn} \rrangle \neq 0$, the vacuum condition leads to $g_3=0$.
  This conflicts the assumption which is written in below (\ref{VC4}).
  Thus, we can say that the second vacuum condition implies $\llangle \CF_{nn} \rrangle = 0$.



\begin{thebibliography}{99}

\bibitem{SeibergWitten1}
  N. Seiberg and E. Witten,
    `` Electric-Magnetic Duality, Monopole Condensation, And Confinement In N=2 Supersymmetric Yang-Mills Theory,''
  Nucl.\ Phys.\ B {\bf 426} (1994) 19, [arXiv:hep-th/9407087]. 
  
\bibitem{SeibergWitten2} 
  N. Seiberg and E. Witten,
    ``Monopoles, Duality and Chiral Symmetry Breaking in N=2 Supersymmetric QCD,''
  Nucl.\ Phys.\ B {\bf 431} (1994) 484, [arXiv:hep-th/9408099].
  
\bibitem{Castellani}
  L. Castellani, R. D'Auria and S. Ferrara
    ``Special K\"ahler Geometry: An Intrinsic Formulation from N=2 Space-time Supersymmetry,''
  Phys.\ Lett.\ B {\bf 241} (1990) 57.

\bibitem{Strominger}
  A. Strominger,
    ``Special Geometry,''
  Comm. Math. Phys. 133 (1990) 163.

\bibitem{D'Auria}
  R.~D'Auria, S.~Ferrara and P.~Fre,
    ``Special And Quaternionic Isometries: General Couplings In N=2 Supergravity And The Scalar Potential,''
  Nucl.\ Phys.\ B {\bf 359} (1991) 705.

\bibitem{Andrianopoli0}
  L. Andrianopoli, M. Bertolini, A. Ceresole, R. D'Auria, S. Ferrara and P. Fre,
    ``General Matter Coupled N=2 Supergravity,''
  Nucl.\ Phys.\ B {\bf 476} (1996) 397, [arXiv:hep-th/9603004].

\bibitem{Andrianopoli}
  L.~Andrianopoli, M.~Bertolini, A.~Ceresole, R.~D'Auria, S.~Ferrara, P.~Fre and T.~Magri,
    ``N = 2 supergravity and N = 2 super Yang-Mills theory on general scalar
       manifolds: Symplectic covariance, gaugings and the momentum map,''
  J.\ Geom.\ Phys.\  {\bf 23} (1997) 111, [arXiv:hep-th/9605032].

\bibitem{VanPro}
  B. Craps, F. Roose, W. Troost and A. Van Proeyen,
    ``What is Special K\"ahler Geometry ?,''
  Nucl.\ Phys.\ B {\bf 503} (1997) 565, [arXiv:hep-th/9703082].
  
\bibitem{VanPro2}
  M. Billo, F. Denef, P. Fre, I. Pesando, W. Troost, A. Van Proeyen and D. Zanon,
    ``The rigid limit in Special Kahler geometry; From K3-fibrations to Special Riemann surfaces: a detailed case study,''
  Class. Quant. Grav. {\bf 15} (1998) 2083, [arXiv:hep-th/9803228].
  
\bibitem{VanPro3}
  P. Claus, K. Van Hoof and A. Van Proeyen,
    ``A symplectic covariant formulation of special Kahler geometry in superconformal calculus,''
  Class. Quant. Grav. {\bf 16} (1999) 2625, [arXiv:hep-th/9904066].

\bibitem{Cecotti2}
  S. Cecotti, L. Girardello and M. Porrati,
    `` An Exceptional N=2 Supergravity with Flat Potential and Partial SuperHiggs,''
  Phys.\ Lett.\ B {\bf 168} (1986) 83.

\bibitem{Ferrara1}
  S.~Ferrara, L.~Girardello and M.~Porrati,
    ``Minimal Higgs Branch for the Breaking of Half of the Supersymmetries in N=2 Supergravity,''
  Phys.\ Lett.\ B {\bf 366} (1996) 155, [arXiv:hep-th/9510074].

\bibitem{Ferrara2}
  S.~Ferrara, L.~Girardello and M.~Porrati,
    ``Spontaneous Breaking of N=2 to N=1 in Rigid and Local Supersymmetric Theories,''
  Phys.\ Lett.\ B {\bf 376} (1996) 275, [arXiv:hep-th/9512180].

\bibitem{APT}
  I.~Antoniadis, H.~Partouche and T.R.~Taylor, 
    ``Spontaneous Breaking of N=2 Global Supersymmetry,"
  Phys.\ Lett.\ B {\bf 372} (1996) 83, [arXiv:hep-th/9512006].
  
\bibitem{APT2}
  I. Antoniadis and T.R. Taylor
    ``Dual N=2 SUSY Breaking,''
  Fortsch. Phys. {\bf 44} (1996) 487, [arXiv:hep-th/9604062].
  
\bibitem{APT3}
  H. Partouche and B. Pioline,
    ``Partial Spontaneous Breaking of Global Supersymmetry,''
  Nucl. Phys. Proc. Suppl. {\bf 56B} (1997) 322, [arXiv:hep-th/9702115].

\bibitem{Fre}
   P.~Fre, L.~Girardello, I.~Pesando and M.~Trigiante,
    ``Spontaneous N=2 $\rightarrow$ N=1 local supersymmetry breaking with surviving local gauge group,''
   Nucl.\ Phys.\ B {\bf 493} (1997) 231, [arXiv:hep-th/9607032].

\bibitem{Porrati}
  M. Porrati,
    ``Spontaneous Breaking of Extended Supersymmetry in Global and Local Theories,''
  Nucl. Phys. Proc. Suppl. {\bf 55B} (1997) 240, [arXiv:hep-th/9609073].

\bibitem{Louis}
  J. Louis
    ``Aspects of Spontaneous N=2 $\rightarrow$ N=1 Supersymmetry Breaking in Supergravity,''
  [arXiv:hep-th/0203138].

\bibitem{Andrianopoli1}
  L. Andrianopoli, R. D'Auria, S. Ferrara and M.A. Lledo,
    ``Duality and Spontaneously Broken Supergravity in Flat Backgrounds,''
  Nucl.\ Phys.\ B {\bf 640} (2002) 63, [arXiv:hep-th/0204145].

\bibitem{Andrianopoli2}
  L. Andrianopoli, R. D'Auria, S. Ferrara and M.A. Lledo,
    ``N=2 Super-Higgs, N=1 Poincare Vacua and Quaternionic Geometry,''
  JHEP {\bf 0301} (2003) 045, [arXiv:hep-th/0212236].

\bibitem{FIS12}
  K.~Fujiwara, H.~Itoyama and M.~Sakaguchi,
    ``Supersymmetric U(N) Gauge Model and Partial Breaking of N=2 Supersymmetry,''
  Prog. Theor. Phys. {\bf 113} (2005) 429,  [arXiv:hep-th/0409060].
  
\bibitem{FIS13}
  K.~Fujiwara, H.~Itoyama and M.~Sakaguchi,
    ``U(N) Gauge Model and Partial Breaking of N=2 Supersymmetry,''
  [arXiv:hep-th/0410132].

\bibitem{FIS3}
  K.~Fujiwara, H.~Itoyama and M.~Sakaguchi,
    ``Partial Breaking of N=2 Supersymmetry and of Gauge Symmetry in the U(N) Gauge Model,''
  Nucl.\ Phys.\ B {\bf 723} (2005) 33, [arXiv:hep-th/0503113].

\bibitem{FIS4}
  K.~Fujiwara, H.~Itoyama and M.~Sakaguchi,
    ``Partial Supersymmetry Breaking and N=2 U(N(C)) Gauge Model with Hypermultiplets in Harmonic Superspace,''
  [arXiv:hep-th/0510255]

\bibitem{David}
  J.R.~David, E.~Gava and K.S.~Narain, 
    `` Partial N = 2 $ \rightarrow $ N = 1 supersymmetry breaking and gravity deformed chiral rings,''
  JHEP {\bf 0406} (2004) 041, [arXiv:hep-th/0311086].

\bibitem{Cecotti}
  S. Cecotti, L. Girardello and M. Porrati,
    ``Constraints on Partial Super-Higgs,''
  Nucl.\ Phys.\ B {\bf 268} (1986) 295.
  
\bibitem{WB}
  J. Wess and J. Bagger
    ``Supersymmetry and Supergravity,''
  (second edition, Princeton University Press, 1992).

\bibitem{Andri1}
  L. Andrianopoli, R. D'Auria and S. Ferrara,
    ``Supersymmetry reduction of N-extended supergravities in four dimensions,''
  JHEP {\bf 0203} (2002) 025, [arXiv:hep-th/0110277].

\bibitem{Andri2}
  L. Andrianopoli, R. D'Auria and S. Ferrara,
    ``Consistent reduction of N=2 $\rightarrow$ N=1 four dimensional supergravity coupled to matter,''
  Nucl.\ Phys.\ B {\bf 628} (2002) 387, [arXiv:hep-th/0112192].
  
\end{thebibliography}
\end{document}